\documentclass[amsmath,amssymb,aps,prd,twocolumn,superscriptaddress,10pt,nofootinbib]{revtex4-2}

\usepackage{graphicx}
\usepackage{dcolumn}
\usepackage{subfigure}
\usepackage{color}
\usepackage{url}
\usepackage{rotating}
\usepackage{xspace}
\usepackage[dvipsnames]{xcolor}
\usepackage[colorlinks,urlcolor=Blue,citecolor=Blue,linkcolor=Blue,pdfusetitle]{hyperref}

\newcommand\apjl{The Astrophysical Journal Letters}

\newcommand\mnras{Monthly Notices of the Royal Astronomical Society}
\newcommand\aap{Astronomy and Astrophysics}

\newcommand\bain{Bulletin of the Astronomical Institutes of the Netherlands}

\begin{document}

\title{Inferring binary parameters with dual-line gravitational wave detection from tight inspiraling double neutron stars}

\author{Wen-Fan Feng}
\affiliation{MOE Key Laboratory of Fundamental Physical Quantities Measurements, Hubei Key Laboratory of Gravitation and Quantum Physics, PGMF, Department of Astronomy and School of Physics, Huazhong University of Science and Technology, Wuhan 430074, China}

\author{Jie-Wen Chen}
\affiliation{National Time Service Center, Chinese Academy of Sciences, Xi’an 710600, China}
\affiliation{Key Laboratory of Time and Frequency Primary Standards, Chinese Academy of Sciences, Xi’an 710600, China}

\author{Tan Liu}
\email{lewton@mail.ustc.edu.cn}
\affiliation{MOE Key Laboratory of Fundamental Physical Quantities Measurements, Hubei Key Laboratory of Gravitation and Quantum Physics, PGMF, Department of Astronomy and School of Physics, Huazhong University of Science and Technology, Wuhan 430074, China}
\affiliation{School of Fundamental Physics and Mathematical Sciences, Hangzhou Institute for Advanced Study, UCAS, Hangzhou 310024, China}
\affiliation{University of Chinese Academy of Sciences, 100049/100190 Beijing, China}

\author{Yan Wang}
\email{ywang12@hust.edu.cn}
\affiliation{MOE Key Laboratory of Fundamental Physical Quantities Measurements, Hubei Key Laboratory of Gravitation and Quantum Physics, PGMF, Department of Astronomy and School of Physics, Huazhong University of Science and Technology, Wuhan 430074, China} 

\author{Soumya D.~Mohanty}
\affiliation{Department of Physics and Astronomy, University of Texas Rio Grande Valley, 
Brownsville, Texas 78520, USA}
\affiliation{Department of Physics, IIT Hyderabad, Kandai, Telangana-502284, India}

\date{\today}

\begin{abstract}

Neutron star (NS) binaries can be potentially intriguing gravitational wave (GW) sources, with both high- and low-frequency radiations from the possibly aspherical individual stars and the binary orbit, respectively. The successful detection of such a dual-line source could provide fresh insights into binary geometry and NS physics. 
In the absence of electromagnetic observations, we develop a strategy for inferring the spin-orbit misalignment angle using the tight dual-line double NS system under the spin-orbit coupling. 
Based on the four-year joint detection of a typical dual-line system with LISA and Cosmic Explorer, we find that the misalignment angle and the NS moment of inertia can be measured with sub-percentage and 5\% accuracy, respectively. 

\end{abstract}

\maketitle

\section{Introduction}
Future space-borne detectors aimed at low-frequency gravitational wave (GW) signals, such as LISA \cite{LISA2017} and TianQin \cite{TianQin2016}, are expected to be operational in the 2030s, almost simultaneously with the next generation of ground-based GW observatories targeting high-frequency GWs, such as Einstein Telescope (ET) \cite{ET2010} and Cosmic Explorer (CE) \cite{CosmicExplorer2022}. This opens up the exciting possibility of successful synergistic observations of sources that emit both low- and high-frequency GWs. One such promising class of sources is binary systems containing at least one aspherical neutron star (NS). Such a system would radiate nearly monochromatic GWs at both low frequencies due to orbital motion as well as at high frequencies due to the spins of the NS components.

Previous studies about dual-line systems have focused on the ultracompact X-ray binaries (UCXBs), specializing in the NS-white dwarf (WD) binaries \cite{Tauris2018, Chen2021, Suvorov2021}. 
The numerical simulation of binary stellar evolution has shown that the low-frequency GWs emitted by the binary system consisting of a $1.6~M_{\odot}$ NS and a $0.16~M_{\odot}$ helium WD can enter the LISA band \cite{Tauris2018}. 
Due to the narrow mass range of the WD, this allows the precise measurements of the NS mass to the percent level. By measuring the dual-line strain amplitudes, the combination of NS moment of inertia and the equatorial ellipticity can be constrained \cite{Tauris2018}. 
Furthermore, with the help of the pulsar timing observations of the binary pulsar, the NS moment of inertia and the equatorial ellipticity can be derived separately, assuming that the spin-down effect is dominated by the gravitational radiation rather than the electromagnetic radiation \cite{Chen2021}.
From a practical detection point of view, the dual-line detectability of twelve known UCXBs with sub-hour orbital periods was evaluated based on the GW strains from different radiation mechanisms \cite{Suvorov2021}.

In this work, we focus on tight double NS (DNS) systems in the mHz band for LISA or TianQin. 
In a previous paper  \cite{Feng2023b}, we have improved the modeling of the dual-line GW radiation, especially for high-frequency GW radiation from a triaxial nonaligned NS in the binary, by incorporating the spin-orbit coupling effect. These waveforms contain information about binary geometry and NS physics.
Using only the joint dual-line GW detection without the aid of electromagnetic observations, we develop
a strategy to measure the spin-orbit misalignment angle\footnote{Since the orbital angular momentum $\boldsymbol{L}$ is approximately aligned with $\boldsymbol{J}$ with a negligible deviation of $\sim 10^{-4}~{\rm rad}$ for a DNS in mHz band, we directly take $\theta_S$ \cite{Feng2023b} between the spin $\boldsymbol{S}$ and $\boldsymbol{J}$ as the spin-orbit misalignment angle \cite{Clifton2008}.} (or its supplementary angle) and the NS moment of inertia with sub-percentage and percentage accuracy, respectively.

Our work is expected to have significant implications for the astrophysical results of the inspiraling DNSs. 
From the perspective of compact binary star evolution, the spin-orbit misalignment angle may carry crucial information for constraining various supernova formation channels~\cite{1961BAN....15..265B,Vigna-Gómez_2019,2019MNRAS.487.1178P,Chattopadhyay2020}, since this misalignment can arise from the kick velocity imparted to the neutron star at birth as a result of asymmetry in the associated supernova explosion~\cite{Lai2003,lr2014}.
However, LISA or TianQin cannot strictly constrain this angle since the spin-orbit coupling effect is better measured during binary merging \cite{Poisson1995}.
Besides, this misalignment angle suffers from considerable uncertainties from radio observations \cite{Tauris2017}.
For NS moment of inertia, PSR J0737-3039A is anticipated to reach the accuracy of $\sim 10\%$ only by measuring the periastron advance through long-term timing observation \cite{Lattimer2005, Hu2020MNRAS}.
Although the GW observation of the merging DNS \cite{Landry2018} and its combination with the mass-radius determination from the X-ray emission at the NS surface \cite{Bogdanov2019} have been used to infer the moment of inertia, it is still subject to large uncertainties \cite{Silva2021}.
Here, we propose a complementary method to infer these two parameters with higher measurement accuracy using only the GW observation.

\section{Number of tight Galactic DNSs}
Binary population synthesis simulations \cite{Nelemans2001, Belczynski2010, Liu2014, Lau2020, Breivik2020, Wagg2022} (or simplified backward evolution simulations \cite{Andrews2020, Feng2023}) for Galactic DNSs have made predictions for the number of DNSs detectable by either LISA or TianQin.
These detectable DNSs are generally spread over a relatively wide GW frequency band, i.e. $\sim 0.1-10~{\rm mHz}$.
At frequencies of $\lesssim 3~{\rm mHz}$, detections of these DNSs are limited by test mass acceleration noise and galactic confusion noise \cite{Robson2019}.
However, at frequencies of $\gtrsim 3~{\rm mHz}$, these sources show a higher signal-to-noise ratio $\rho$ because the strain amplitude $\propto f_{\rm g}^{2/3}$ with $f_{\rm g}$ the GW frequency.
Therefore, the number in the frequency band of $\gtrsim 3~{\rm mHz}$ can be approximated as a detectable number.
The eccentricities of these DNSs are assumed to be neglected since the most probable eccentricity of tight LISA DNSs is about 0.01 for orbital periods shorter than $10~{\rm min}$ according to binary population simulations \cite{Wagg2022}. 

The number density distribution over the frequency bin for compact binary systems can be calculated with GW frequency derivative $\dot{f}_{\rm{g}} \equiv df_{\rm{g}}/dt$ and merger rate $R$ as $dN/df_{\rm{g}} = R/\dot{f}_{\rm{g}}$ \cite{Kyutoku2016}. The detectable number of the tight Galactic DNSs for LISA or TianQin can be obtained by integrating the above number density as \cite{Kyutoku2016}
%
\begin{equation} 
N(>f_{\rm{g}})= 6\left(\frac{1.2M_{\odot}}{\mathcal{M}}\right)^{5/3}\left(\frac{3.3\mathrm{mHz}}{f_{\rm{g}}}\right)^{8/3} \left(\frac{R_{\mathrm{G}}}{100/\mathrm{Myr}}\right) \,.
\end{equation}
%
The fiducial value of chirp mass $\mathcal{M}= (m_1 m_2)^{3/5}/(m_1+m_2)^{ 1/5} = 1.2 ~M_\odot$ corresponds to component masses of $m_1 = m_2 = 1.4~M_{\odot}$.
The GW frequency of $f_{\rm g}=3.3~{\rm mHz}$ corresponds to a circular binary orbital period of $P_{\rm b}=10~{\rm min}$.
The local volumetric merger rate of the DNS varies from $10~{\rm Gpc}^{-3}{\rm yr}^{-1}$ up to $10^4~{\rm Gpc}^{-3}{\rm yr}^{-1}$ according to different methods and models (see \cite{Mandel2022} for a review and references therein).
We take the volumetric merger rate for Galactic DNSs to be $1000~{\rm Gpc}^{-3}{\rm yr}^{-1}$ (corresponding to $R_{\rm G} \sim 100~{\rm Myr}^{-1}$) \cite{Burns2020}. 
In this sense, there are about 6 detectable tight DNSs with $P_{\rm b}<10~{\rm min}$ in the Milky Way.

\section{Modulated dual-line GW envelopes by spin-orbit coupling}
We consider a DNS consisting of a spinning NS and a nonspinning companion as in \cite{Feng2023b}. 
For the case of equal mass and ignoring the spin-spin terms, the inspiral and precession equations can be expressed in a simple form (cf. Eqs.~(40) of \cite{Apostolatos1994}).
Let $\boldsymbol{S}$, $\boldsymbol{L}$, and $\boldsymbol{J}$ denote spin, orbital, and total angular momentum, respectively.
For simple precession, the unit vector $\hat{\boldsymbol{S}}$ and $\hat{\boldsymbol{L}}$ both precess around the unit vector $\hat{\boldsymbol{J}}$ with an angular velocity 
\begin{equation}\label{eq_Omegapre}
\Omega_{\mathrm{pre}}=\frac{G J}{c^2 r^3}\left(2+\frac{3 m_2}{2 m_1}\right) \,.
\end{equation}
Although the radiation-reaction-induced orbital shrinkage causes the magnitude $L$ (or $J$) of $\boldsymbol{L}$ (or $\boldsymbol{J}$) to decrease, for a typical tight DNS with a merger time $\sim O(10^4~{\rm yr})$ and a four-year observation time (see Sec.~\ref{sec_infermisalignangle}), 
the relative variation of $L$, $J$, and ${\Omega}_{\rm pre}$ are all $\lesssim 10^{-5}$, so we can assume that they remain approximately constant in the analysis.

\begin{figure}[!htbp]
\centering
\includegraphics[scale=0.5]{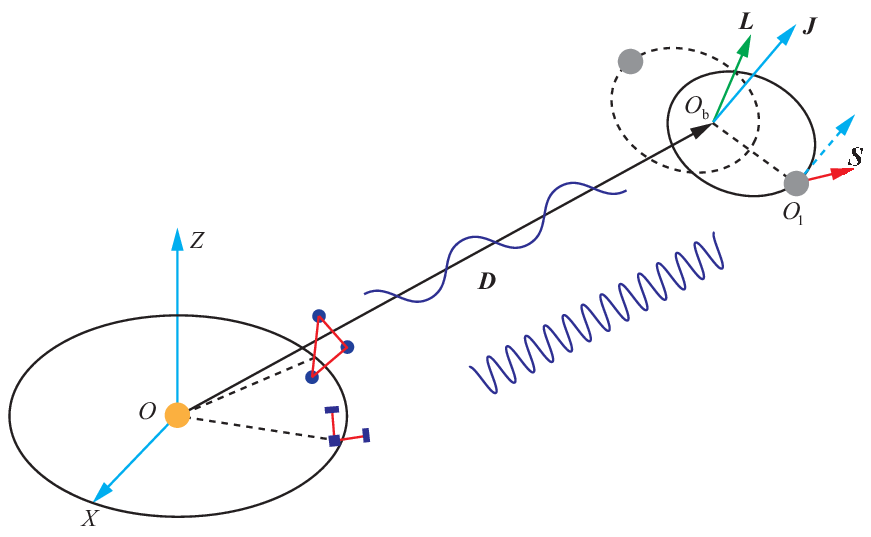}
\caption{Schematic of a dual-line GW source and its detection. }
\label{fig_dual-line}
\end{figure}

Since $\boldsymbol{L}$ is associated with low-frequency GWs from the orbital motion of the DNS and $\boldsymbol{S}$ is related to high-frequency GWs from the spinning NS, the spin-orbit coupling strengthens the link between the dual-line GWs. 
In the $J$-aligned coordinate system \cite{Feng2023b}, constructed with $Z_J$ axis aligned with $\boldsymbol J$, we assume that $\theta_L$ ($\theta_S$) is the angle between $\boldsymbol{L}$ ($\boldsymbol{S}$) and $\boldsymbol{J}$ and $\iota$ is the binary inclination with respect to $\boldsymbol{J}$.
The precession equation of the binary orbit
%
${d \hat{\boldsymbol{L}}}/{d t}= \boldsymbol{\Omega}_{\rm pre} \times \hat{\boldsymbol{L}} $
can be solved with the initial condition $\hat{\boldsymbol{L}}(t=0)=(0,-\sin{\theta_L},\cos{\theta_L})$ as
%
$\hat{\boldsymbol{L}}=(\sin{\theta_L}\sin(t\Omega_{\rm pre}),-\sin{\theta_L}\cos(t\Omega_{\rm pre}),\cos{\theta_L}) \,.$
%
And since the unit vector pointing towards the solar system barycenter is $\hat{\boldsymbol{D}} = (0,\sin{\iota},\cos{\iota})$, it follows that
$\hat{\boldsymbol{L}} \cdot \hat{\boldsymbol{D}} = \cos{\theta_L} \cos{\iota} - \cos(t\Omega_{\rm pre}) \sin{\theta_L}\sin{\iota} \,.$
%
The low-frequency GW amplitude envelopes radiated by the binary orbital motion under the spin-orbit coupling are (we only take positive values, cf. Eqs.~(18) of \cite{Apostolatos1994})
\begin{subequations}\label{eq_orbenvelope}
\begin{align}
a_{2+}^{\rm orb} &= \frac{h_0}{2} \left[ 1+(\cos{\theta_L} \cos{\iota} - \cos(t\Omega_{\rm pre}) \sin{\theta_L}\sin{\iota})^2  \right] \,, \\
a_{2\times}^{\rm orb} &= h_0 \left[\cos{\theta_L} \cos{\iota} - \cos(t\Omega_{\rm pre}) \sin{\theta_L}\sin{\iota} \right] \,,
\end{align}
\end{subequations}
where $h_0 \equiv \frac{4(G \mathcal{M})^{5/3}(\pi f_{\rm g})^{2/3}}{c^4 D}$, $D$ is the distance from the solar system barycenter to the binary barycenter. 
The high-frequency GW amplitude envelopes of the triaxial nonaligned NS in the binary under the spin-orbit coupling are (we only take positive values, cf. Eqs. (30) in \cite{Feng2023b}) 
%
\begin{subequations}\label{eq_sphenvelope}
\begin{align}\nonumber
a_{1+}^{\rm sp} =&~ \frac{h_{10}}{8}\Big[\Big(4\cos (2{\theta _S})\cos (t{\Omega _{{\rm{pre}}}})\sin (2\iota )  \\ \nonumber
&+ \sin (2{\theta _S})[ 6{\sin ^2}\iota - (3 + \cos (2\iota )) \cos (2t{\Omega _{{\rm{pre}}}})] \Big)^2 \\ \nonumber
&+ 4\Big( - 2\cos {\theta _S}\sin (2\iota )\sin (t{\Omega _{{\rm{pre}}}}) \\ 
&+ (3 + \cos (2\iota ))\sin {\theta _S}\sin (2t{\Omega _{{\rm{pre}}}})\Big)^2\Big]^{1/2} \,, \\ \nonumber
a_{1\times} ^{\rm sp} =&~ \frac{h_{10}}{2}\Big[\Big(2\sin \iota \cos (2{\theta _S})\sin (t{\Omega _{{\rm{pre}}}}) \\ \nonumber
&- \cos \iota \sin(2{\theta _S})\sin (2t{\Omega _{{\rm{pre}}}})\Big)^2 \\ \nonumber
&+ 4\Big(\sin \iota \cos {\theta _S}\cos (t{\Omega _{{\rm{pre}}}}) \\ 
&- \cos \iota \cos (2t{\Omega _{{\rm{pre}}}})\sin {\theta _S}\Big)^2\Big]^{1/2}  \,, \\ \nonumber
a_{2+}^{\rm sp} =&~ \frac{h_{20}}{4}\Big[\Big(- 2\sin {\theta _S}\sin (2\iota )\sin (t{\Omega _{{\rm{pre}}}}) \\ \nonumber
&-\cos \theta_S (3 + \cos (2\iota ))\sin (2t{\Omega _{\rm{pre}}}) \Big)^2 \\ \nonumber
&\Big((\sin^4  {\frac{\theta _S}{2}} +\cos^4  {\frac{\theta _S}{2}} ) (3 + \cos (2\iota )) \cos (2t{\Omega _{\rm{pre}}})    \\ 
&+ 3{\sin ^2}{\theta _S}{\sin ^2}\iota + \cos (t{\Omega _{{\rm{pre}}}}) \sin(2{\theta _S})\sin (2\iota )\Big)^2  \Big]^{1/2}  \,, \\ \nonumber
a_{2\times} ^{\rm sp} =&~ \frac{h_{20}}{4}\Big[\Big(- 2\sin \iota \sin (2{\theta _S})\sin (t{\Omega _{{\rm{pre}}}}) \\ \nonumber
&- \cos \iota (3 + \cos (2{\theta _S}))\sin (2t{\Omega _{{\rm{pre}}}})\Big)^2 \\ \nonumber
&+ 16\Big(\cos \iota \cos {\theta _S}\cos (2t{\Omega _{{\rm{pre}}}}) \\ 
&+ \sin \iota \cos (t{\Omega _{{\rm{pre}}}})\sin {\theta _S}\Big)^2\Big]^{1/2}  \,,
\end{align}
\end{subequations}
%
where $h_{10} \equiv \frac{2G{b^2}{I_3}\epsilon \gamma }{{c^4}D}, h_{20} \equiv \frac{64G{b^2}{I_3}\epsilon \kappa }{{c^4}D}$, $b$ is the initial angular frequency of the spinning NS along one of the principal axes of inertia (corresponding to the principal moment of inertia $I_3$) in the body frame, parameters $\epsilon$, $\gamma$, and $\kappa$ characterize the oblateness, wobble angle, and non-axisymmetry of the NS, respectively \cite{Broeck2005, Gao2020}. The amplitude envelopes of $a_{3+,\times}^{\rm sp}$ are similar to $a_{2+,\times}^{\rm sp}$ except that $h_{20}$ is replaced by $h_{30} \equiv \frac{4 G{b^2}{I_3}\epsilon \gamma^2}{{c^4}D}$.
We have neglected the variation in the frequency of the NS spin over several years of observations due to the long spin-down timescale ($\sim 500-1000~{\rm Myr}$) \cite{Oslowski2011, Chattopadhyay2020, Chattopadhyay2021}.

\section{Inferring binary parameters}

Observations of continuous wave signals from the dual-line GW source span months or years, resulting in the accumulation of substantial datasets. However, since these signals are confined to an extremely narrow frequency band, the application of the complex heterodyne technique allows us to significantly reduce the size of the dataset by incorporating precise phase details obtained from observations in radio or X-ray \cite{Dupuis2005, Niebauer1993}.
In this case, the amplitudes of the GW polarizations can be reconstructed from the heterodyned and down-sampled noisy detector strain data $B(t_k)$ by minimizing the $\chi^2$ of the system \cite{Isi2015}. For template $T(t_k)$, that is
\begin{equation}\label{eq_chi2}
\chi^2=\sum_{k=0}^{N-1}\left[\frac{B(t_k)-T(t_k)}{\sigma_k}\right]^2 \,,
\end{equation}
where $\sigma_k$ is the measurement uncertainty of the $k$th data point at time $t_k$. 
General relativity predicts only two states of polarization $p=+, \times$, the relative amplitude $a_p$ and phase $\phi_p$ coefficients in $B(t_k)$ are determined by the physical model. Choose the antenna pattern functions $F_p(t)$ as the basis functions and polarization prefactors $H_p=a_p {\rm exp}(i\phi_p)$ as the coefficients those are fitted for, then the template can be constructed as $T(t_k)=\frac{1}{2}\sum_p H_p F_p(t_k)$. The signal-to-noise ratio of the resulting fit is \cite{Isi2015}
\begin{equation}\label{eq_snr_fit}
\rho = \sqrt{H^{\dagger}C^{-1}H} \,,
\end{equation}
where $\dagger$ denotes Hermitian conjugation, the vector $H=\{H_{+}, H_{\times}\}$ is the solution by minimizing Eq.~(\ref{eq_chi2}), the covariance matrix $C=(A^T A)^{-1}$ can be computed by the design matrix $A_{kp}=F_p(t_k)/\sigma_k$, and the variance of the estimated $H_j$ can be found as $\Delta^2 H_j=C_{jj}$ \cite{Numericalrecipes}. 

In the absence of electromagnetic observations, to know the phase, we can perform the following strategy of joint detection: LISA first measures the parameters of the tight DNS with high accuracies of sky localization $\Delta \Omega \sim 10^{-3}~{\rm deg}^2$, distance $\Delta D/D \sim 10^{-3}$, orbital period $\Delta P_{\rm b}/P_{\rm b} \sim 10^{-9}$ for $\rho \sim O(10^3)$ \cite{Feng2023}, then CE or ET conducts a directed search for continuous GWs from the NS in the binary \cite{Sammut2014, Zhang2021}.

We first derive the relative reconstruction error for the case of constant polarization amplitude. For general case, $\sum_{k} A_{k+}^2 \approx \sum_{k} A_{k\times}^2$ and $\sum_{k} A_{k+}^2 \sum_{k} A_{k\times}^2 \gg \left(\sum_{k} A_{k+} A_{k\times} \right)^2$, so that $\Delta^2 H_{p} \approx 1/ \sum_{k} A_{kp}^2$, where $\sum_k$ stands for $\sum_{k=0}^{N-1}$.
If $a_{+} \approx a_{\times}$, then the relative $1\sigma$ uncertainties of the recovered polarization amplitudes are 
\begin{align}\label{eq_invrho}
\frac{\Delta a_{+}}{a_{+}} \approx 
\frac{\Delta a_{\times}}{a_{\times}} &\approx \frac{1}{\rho} \,.
\end{align}

We shall justify that this relationship also holds approximately for time-varying amplitude envelopes [cf. Eqs.~(\ref{eq_orbenvelope}) and (\ref{eq_sphenvelope})]. 
Based on the low-frequency radiation from the binary inspiral, LISA can measure the chirp mass \cite{Feng2023} and the total mass \cite{Seto2001} to an accuracy of $\sim 10^{-4}$ for $\rho \sim O(10^3)$, which derives the measurement accuracies of the component masses and the relativistic precession velocity $\Omega_{\rm pre}$ [cf. Eq.~(\ref{eq_Omegapre})] to be $\sim 10^{-4}$.
As we will see below, all quantities of interest have a measurement accuracy greater than $\gtrsim 10^{-3}$, so we ignore the accuracy of $\sim 10^{-4}$.
Once $\Omega_{\rm pre}$ has been accurately measured, the modulated envelopes under the spin-orbit coupling can be expanded as $a_p(t)=\sum_{n=0}^{E-1}a_p^n \cos(n\Omega_{\rm pre}t)$, where $E$ is the order of trigonometric expansion. The new set of basis formed by combining this basis $\{1,  \cos(n\Omega_{\rm pre}t)\}$ with the sidereal basis $\{1, \cos(m\Omega_{\rm E}t), \sin(m\Omega_{\rm E}t)\}$ in $F_p(t)$ \cite{Jaranowski1999PhRvD} can be approximated as an orthogonal basis. In this case, $a_{p}^n$ is the coefficient that is fitted for by the above data processing procedure, and its variance $\Delta ^2 a_p^n \approx (a_p^n/\rho)^2$. Finally, the error propagation shows that $\Delta ^2 a_p(t)/a^2_p(t)=\sum_{n=0}^{E-1}\Delta ^2 a_p^n \cos^2(n\Omega_{\rm pre}t)/a^2_p(t) \approx 1/\rho^2$.
Therefore, we can reasonably anticipate that the polarization amplitude envelopes can be extracted from the dual-line GW signals, and the relative error in their reconstruction is approximately inversely proportional to the signal-to-noise ratio.

\subsection{Spin-orbit misalignment angle}
\label{sec_infermisalignangle}
By utilizing these reconstructed amplitude envelopes, we can employ them to infer the binary parameters according to Eqs.~(\ref{eq_orbenvelope}) and (\ref{eq_sphenvelope}).
To avoid possible large errors at a single reconstruction point \cite{Kuwahara2022}, we average the amplitude envelopes over a precession period as follows,
\begin{align} 
\left \langle a \right \rangle = \frac{1}{T_{\rm pre}} \int_{0}^{T_{\rm pre}} a ~dt  \,.
\end{align}
The error introduced by the precession period $T_{\rm pre}=2\pi/{\Omega_{\rm pre}}$ with an accuracy of $\sim 10^{-4}$ (cf. the discussion below Eq.~(\ref{eq_invrho})) is also ignored.
Assuming the reconstructed $\left \langle a \right \rangle$ lies within the range of true value $\pm 1\sigma$ uncertainty.

For a typical DNS with parameters as in \cite{Feng2023b}, $m_1 = m_2 = 1.4~M_{\odot}, P_{\rm b}=10~{\rm min}, D=1~{\rm kpc}, \iota=\pi/4$.
The spinning NS in the DNS is parameterized by $P_{\rm s}=10~{\rm ms}, I_3=2.0\times 10^{38}~{\rm kg~m^2}, \epsilon=3.6\times 10^{-6}, \kappa=1.75\times 10^{-4}, \gamma=5.0\times 10^{-2}, \theta_{\rm S}=5\pi/12$. Then one can obtain $\Omega_{\rm pre}=2.52\times 10^{-7}~{\rm Hz}$, $T_{\rm pre}=2.50\times 10^{7}~{\rm s}$, $\theta_L=3.7\times 10^{-4}~{\rm rad}$.
Given the source location and polarization angle $(\theta, \phi, \psi)=(0.5,1.5,1.0)~{\rm rad}$ as well as the observation time $T_{\rm obs}=5T_{\rm pre}=1.25\times 10^8~{\rm s}$, LISA can detect the DNS with $\rho_0=2682$, while CE can detect the NS with $\rho_{1}=283$, $\rho_{2}=23$, and $\rho_{3}=21$ for waveform components with $h_{10}$, $h_{20}$, and $h_{30}$, respectively.

\begin{figure*}[!htbp]
\centering
\includegraphics[width=\textwidth]{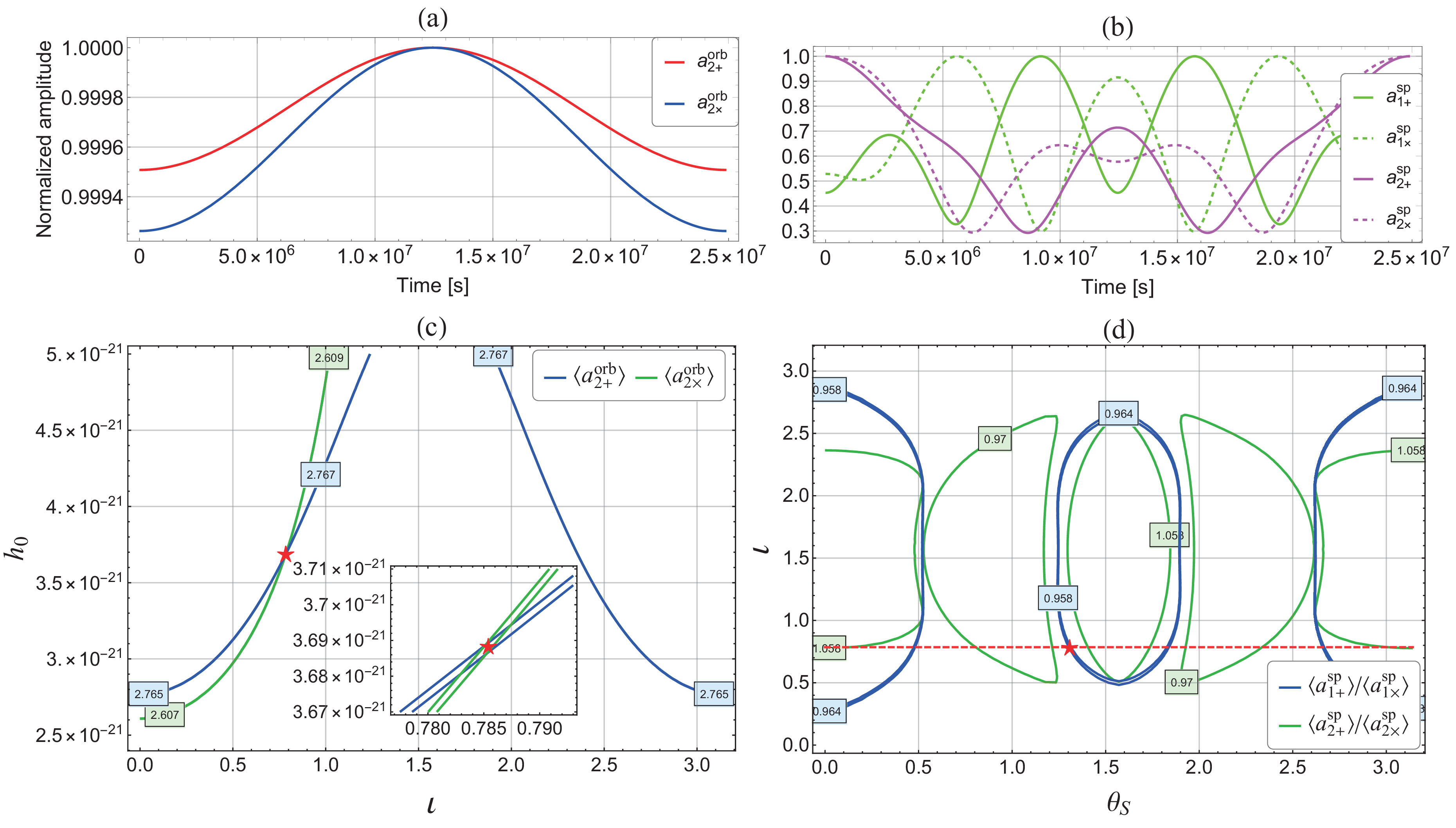}
\caption{Subfigure (a) and (b) depict the normalized ($a/a_{\rm max}$) low- and high-frequency polarization amplitude envelopes from a typical DNS during one precession period $T_{\rm pre}$, with relative changes of $\sim 10^{-3}$ and $100\%$, respectively. The parameter values used here are $m_1 = m_2 = 1.4~M_{\odot}, P_{\rm b}=10~{\rm min}, D=1~{\rm kpc}, \iota=\pi/4, P_{\rm s}=10~{\rm ms}, I_3=2.0\times 10^{38}~{\rm kg~m^2}$, $\epsilon=3.6\times 10^{-6}$, $\kappa=1.75\times 10^{-4}$, $\gamma=5.0\times 10^{-2}, \theta_{\rm S}=5\pi/12$. Subfigure (c) shows the inference for amplitude factor $h_0$ and inclination $\iota$ using low-frequency amplitude envelopes $\left \langle a_{2+}^{\rm orb} \right \rangle$ and $\left \langle a_{2\times}^{\rm orb} \right \rangle$ with $\pm 1\sigma$ uncertainty. In the inset, the overlapped region of these contour lines gives the inferred ranges for $h_0$ and $\iota$ with sub-percentage accuracy, in which the red star denotes their true value. For visualization purposes, the contour labels are $\left \langle a_{2+}^{\rm orb} \right \rangle\times 10^{21}$ and $\left \langle a_{2\times}^{\rm orb} \right \rangle\times 10^{21}$. Subfigure (d) shows the inference for $\theta_S$ and $\iota$ from dual-line detection. The $\iota$ measured (red dashed line) from low-frequency amplitudes in subfigure (c) will greatly constrain the range of $\theta_S$ inferred from $\left \langle a_{1+}^{\rm sp} \right \rangle/\left \langle a_{1\times} ^{\rm sp} \right \rangle$ and $\left \langle a_{2+}^{\rm sp} \right \rangle/\left \langle a_{2\times} ^{\rm sp} \right \rangle$. }
\label{fig_dual-lineinfer}
\end{figure*}

The normalized low- and high-frequency polarization amplitude envelopes from this DNS system are shown in Fig.~\ref{fig_dual-lineinfer}(a) and \ref{fig_dual-lineinfer}(b), with relative changes of $\sim 10^{-3}$ and $100\%$, respectively.

An example of inferring $\iota$ and $h_0$ from low-frequency detection using the average envelopes $\left \langle a_{2+}^{\rm orb} \right \rangle$ and $\left \langle a_{2\times}^{\rm orb} \right \rangle$ with $\pm 1\sigma$ uncertainty is shown in Fig.~\ref{fig_dual-lineinfer}(c). 
In such a system, $L \gg S$, $J \approx L$, 
and  $\cos{2 \theta_L} \simeq \cos{\theta_L} \simeq 1$ with relative errors $\lesssim 10^{-7}$, 
then $\left \langle a_{2+}^{\rm orb} \right \rangle \simeq {h_0}(1+\cos^2\iota)/2$ and $\left \langle a_{2\times}^{\rm orb} \right \rangle \simeq h_0 \cos \iota$ with negligible error. 
The two blue (or green) contours correspond to boundaries of $\pm 1\sigma$ uncertainty, $\left \langle a_{2+}^{\rm orb} \right \rangle (1\pm \rho_0^{-1})$ (or $\left \langle a_{2\times}^{\rm orb} \right \rangle (1\pm \rho_0^{-1})$).
The overlapped region of these lines gives the inferred ranges for $h_0$ and $\iota$ with an accuracy of $0.3\%$ and $0.5\%$, respectively.

The parameter inference by combining dual-line detection from low- and high-frequency is shown in Fig.~\ref{fig_dual-lineinfer}(d).
The relative errors of $\left \langle a_{1+}^{\rm sp} \right \rangle/\left \langle a_{1\times} ^{\rm sp} \right \rangle$ and $\left \langle a_{2+}^{\rm sp} \right \rangle/\left \langle a_{2\times} ^{\rm sp} \right \rangle$ are $\sim 1/{\rho_1}$ and $\sim 1/{\rho_2}$, respectively.
And also plotted are the contours with $\pm 1\sigma$ uncertainty, $(1\pm \rho_1^{-1})\left \langle a_{1+}^{\rm sp} \right \rangle/\left \langle a_{1\times} ^{\rm sp} \right \rangle $ and $(1\pm \rho_2^{-1})\left \langle a_{2+}^{\rm sp} \right \rangle/\left \langle a_{2\times} ^{\rm sp} \right \rangle $.
The overlapped regions of these lines (i.e., the regions enclosed by the blue lines) give the inferred ranges for $\theta_S$ and $\iota$, whose accuracies are poor. When combined with the inferred $\iota$ (red dashed line) from the low-frequency detection, $\theta_S$ can be significantly constrained to an accuracy of $0.8\%$.
Values other than the $\theta_S$ and $\pi-\theta_S$ can be excluded from the amplitude envelopes of the spinning NS [cf. Eqs.~(\ref{eq_sphenvelope})].
Similar to the pulsar observation \cite{Stairs2004}, both $\theta_S$ and $\pi-\theta_S$ yield the same outcomes for precession in General Relativity. Considering that the angular momenta were most likely aligned before the second supernova, the preference is given to the smaller misalignment value based on astrophysical considerations \cite{Bailes1988}.
It should be noted that this degeneracy does not affect the derivation of the amplitude factors below, since the two angles correspond to the same amplitude. 
Our results highlight the importance of synergistic analysis for dual-line observations, as individual analyses at low or high frequencies alone cannot provide a meaningful constraint.

\subsection{Moment of inertia}
\begin{figure}[!htbp]
\centering
\includegraphics[scale=0.2]{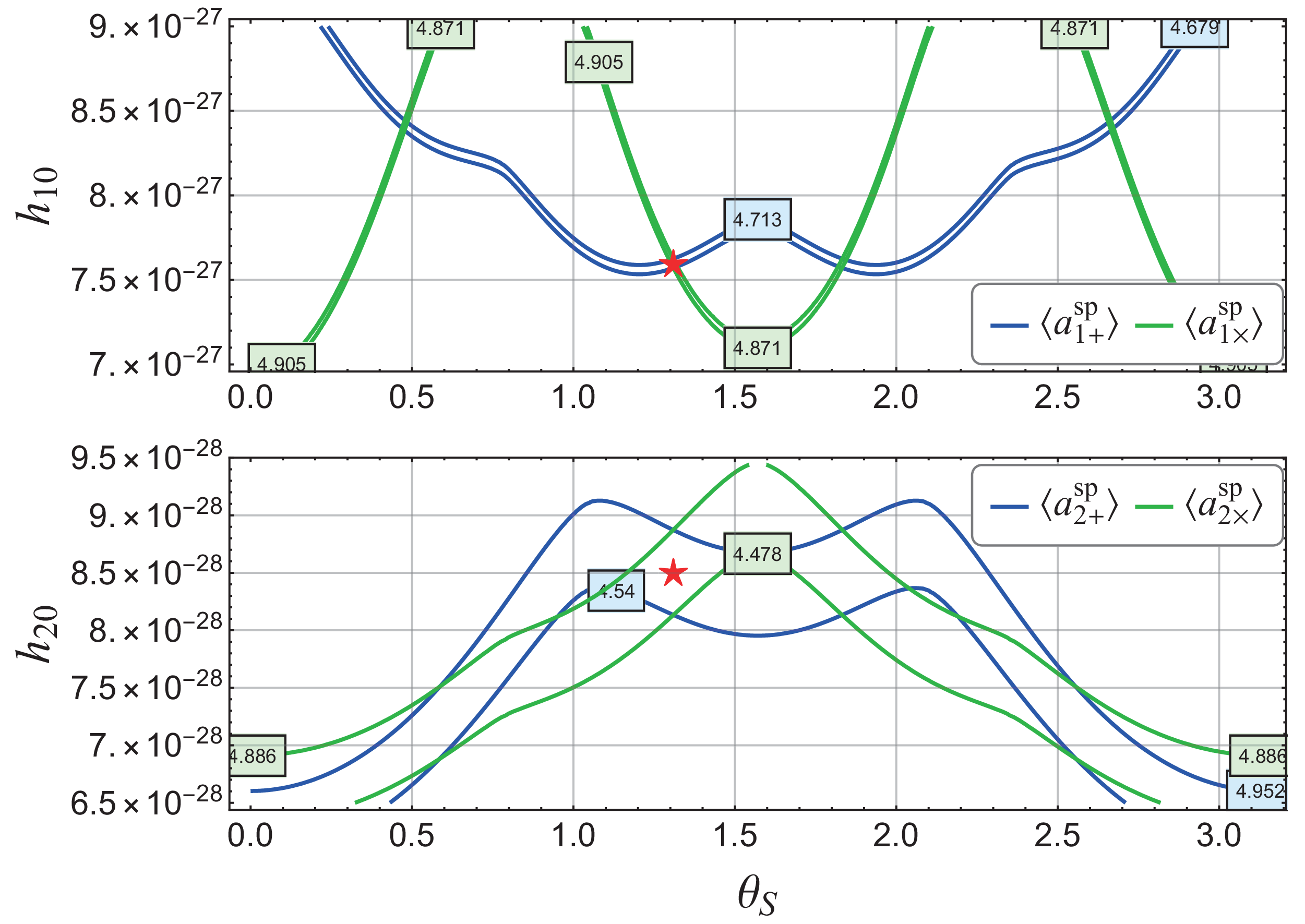}
\caption{Similar to Fig.~\ref{fig_dual-lineinfer}, inference of high-frequency amplitude factors $h_{10}$ and $h_{20}$ with the help of accurate inclination ($\iota$) determination from low-frequency detection. The estimation accuracy of $h_{10}$ and $h_{20}$ is $0.5\%$ and $5\%$, respectively. For visualization purposes, the contour labels are $\left \langle a_{1+}^{\rm sp} \right \rangle\times 10^{27}$, $\left \langle a_{1\times}^{\rm sp} \right \rangle\times 10^{27}$, $\left \langle a_{2+}^{\rm sp} \right \rangle\times 10^{28}$, and $\left \langle a_{2\times}^{\rm sp} \right \rangle\times 10^{28}$.}
\label{fig_h10h20_thetaS}
\end{figure}

By bringing the accurate binary inclination $\iota$ inferred from the low-frequency envelopes into the high-frequency envelopes, utilizing a similar approach in Fig.~\ref{fig_dual-lineinfer}(c) and \ref{fig_dual-lineinfer}(d), we can further infer that the relative errors of the amplitude factors are $\Delta h_{10}/h_{10} \sim 0.5\%$ and $\Delta h_{20}/h_{20} \sim \Delta h_{30}/h_{30} \sim 5\%$ (cf. Fig.~\ref{fig_h10h20_thetaS}).
Since $\gamma=h_{30}/(2h_{10})$ and $\kappa=h_{20}h_{30}/(64h_{10}^2)$, the error propagation shows that $\Delta \gamma/\gamma \sim \Delta \kappa/\kappa \sim 5\%$.
Furthermore, the spectrum analysis of several years of observational data allows us to determine the $\epsilon$ to an accuracy of $\ll 10^{-4}$ with the relation between rotation frequency $\Omega_{\rm r}$ and nonaxisymmetry-induced precession frequency $\Omega_{\rm p}$ of the NS, $\Omega_{\rm p} \simeq \epsilon \Omega_{\rm r}$ (cf. Eq.~(5.7) in \cite{Broeck2005}). This also implies that $b \simeq \Omega_{\rm r} \cos{\gamma}$ can be measured with an accuracy of $\Delta \cos{\gamma}/\cos{\gamma} \sim (\Delta \gamma /\gamma){\gamma}^2 \sim 10^{-4}$.
Finally, with a derived $\Delta D/D \sim 10^{-3}$ from LISA (cf. Eq. (33) in \cite{Feng2023b}), $I_3=\frac{c^4 D}{2G} \frac{h_{10}}{b^2 \epsilon \gamma}$ can be inferred as $\Delta I_3/I_3 \sim \Delta \gamma/\gamma \sim 5\%$. 
This analysis shows that dual-line GW detection is indispensable in inferring NS moment of inertia, in the absence of electromagnetic-assisted observations.

%

\section{Conclusion}
Our study on the modulated waveform structure from the dual-line GW source by incorporating the spin-orbit coupling has unveiled crucial insights into binary geometry and NS physics.
Although multimessenger observations from combining LISA and SKA can also place constraints on the properties of NSs \cite{Thrane2020}, the strategy we use in this paper offers a new measurement possibility in the absence of electromagnetic observations.
With more accurate inferences for the spin-orbit misalignment angle and the NS moment of inertia in the dual-line GW detection, one can expect to place more constraints on the kick in the second supernova explosion in DNSs \cite{Hills1983, Kramer1998, Farr2011} and discriminate between different models of the NS equation of states \cite{Morrison2004, Bejger2005, Lattimer2005, Greif2020, Miao2022}.
Furthermore, the measurement of the spin-orbit misalignment angle could provide clues to distinguish between different SN scenarios with kick velocity \cite{Wong2010, Ferdman2013}. For example, a low spin-orbit misalignment angle measured for PSR J0737-3039A, combined with determined factors such as low pulsar B mass, system eccentricity, and transverse velocity from timing measurements, would strongly support the evidence for the formation of pulsar B through an electron capture supernova event \cite{Kalogera2008}. On the other hand, accretion theory predicts that as pulsar A accretes matter from the progenitor of pulsar B, its spin axis is anticipated to align with the total angular momentum (effectively represented by the orbital angular momentum) of the binary system. In the event of a substantial kick to the system, the resulting misalignment would be equivalent to the angular difference in the orientation of the orbital plane before and after the supernova event \cite{Wex2000}. These rich astrophysical implications of the dual-line systems will be the subject of our future investigations.

\begin{acknowledgments}
We thank Yuanhao Zhang, Jian-Ping Yuan, and Yan-Rong Zhang for helpful discussions on continuous GW searches and radio pulsar observations.
Y.W. gratefully acknowledges support from the National Key Research and Development Program of China (No. 2022YFC2205201 and No. 2020YFC2201400), the National Natural Science Foundation of China (NSFC) under Grants No. 11973024, Major Science and Technology Program of Xinjiang Uygur Autonomous Region (No. 2022A03013-4), and Guangdong Major Project of Basic and Applied Basic Research (Grant No. 2019B030302001). 
S.D.M is supported by U.S. National Science Foundation (NSF) grant PHY-2207935. 
We thank the anonymous referee for helpful comments and suggestions.

\end{acknowledgments}


%

\end{document}